\documentclass[manuscript]{aastex} 
\usepackage{graphicx}
\begin{document}

\title{Transient Fragments in Outbursting Comet 17P/Holmes \thanks{Based on observations obtained with MegaPrime/MegaCam, a joint project of CFHT and CEA/DAPNIA, at the Canada-France-Hawaii Telescope (CFHT) which is operated by the National Research Council (NRC) of Canada, the Institut National des Science de l'Univers of the Centre National de la Recherche Scientifique (CNRS) of France, and the University of Hawaii.} $^{,}$\thanks{This work is based in part on data products produced at the TERAPIX data center located at the Institut d'Astrophysique de Paris. }}
%\date{\today}

\author{Rachel Stevenson}
\affil{Dept. Earth and Space Sciences, UCLA}
\affil{595 Charles Young Drive East, Los Angeles, CA 90095-1567}
\email{stevenson@ucla.edu}
\author{Jan Kleyna}
\affil{Institute for Astronomy, University of Hawaii}
\affil{2680 Woodlawn Drive, Honolulu, HI 96822}
\email{kleyna@ifa.hawaii.edu}
\and
\author{David Jewitt}
\affil{Dept. Earth and Space Sciences, Institute for Geophysics and Planetary Physics, UCLA}
\affil{595 Charles Young Drive East, Los Angeles, CA 90095-1567}
\email{jewitt@ucla.edu}
%\altaffilmark{1, 2, 3}}
%\altaffiltext{1}{Institute for Astronomy, University of Hawaii}
%\altaffiltext{2}{Dept. Earth and Space Sciences, UCLA}
%\altaffiltext{3}{Institute for Geophysics and Planetary Physics}

\begin{abstract}

We present results from a wide-field imaging campaign at the Canada-France-Hawaii Telescope to study the spectacular outburst of comet 17P/Holmes in late 2007.  Using image-processing techniques we probe inside the spherical dust coma and find sixteen fragments having both spatial distribution and kinematics consistent with isotropic ejection from the nucleus.  Photometry of the fragments is inconsistent with scattering from monolithic, inert bodies. Instead, each detected fragment appears to be an active cometesimal producing its own dust coma.  By scaling from the coma of the primary nucleus of 17P/Holmes, assumed to be 1.7 km in radius, we infer that the sixteen fragments have maximum effective radii between $\sim$ 10 m and $\sim$ 100 m on UT 2007 Nov.\ 6.  The fragments subsequently fade at a common rate of $\sim$ 0.2 mag day$^{-1}$, consistent with steady depletion of ices from these bodies in the heat of the Sun.  Our characterization of the fragments supports the hypothesis that a large piece of material broke away from the nucleus and crumbled, expelling smaller, icy shards into the larger dust coma around the nucleus. 

\end{abstract}


 \section{Introduction}

Comet 17P/Holmes is a dynamically and compositionally typical Jupiter Family Comet (Schleicher 2009) but it has exhibited three dramatic outbursts that caused an increase in brightness large enough to lift it from obscurity to naked-eye visibility (Holmes 1892; Palisa 1893; Buzzi et al. 2007).  The first outburst led to its discovery on UT 1892 Nov.\ 6 by Edwin Holmes (Holmes 1892) and was followed by a second outburst three months later in January 1893.  The third outburst, first identified by J.\ A.\ Henriquez Santana on UT 2007 Oct.\ 24 (Buzzi et al. 2007), caused the comet to reach a brightness of 2$^{nd}$ magnitude.  

%be more specific, magnitude changes for first outburst [ref].  could talk about observations from 1st explosion...?

The nature of cometary mass loss varies widely between comets, ranging from gentle outgassing to violent outbursts as observed in the case of 17P/Holmes.  Possible causes of large outbursts are numerous but in this case we are able to rule out several.  The 1892 and 1893 outbursts of 17P/Holmes were attributed to impacts with a satellite (Whipple 1984) but this possibility is rendered extremely unlikely by a third, similar outburst 115 years later.  Rotational breakup requires a rotation period of less than 5.2 hours (assuming a spherical, strengthless body with a density of 400 kg m$^{-3}$; Richardson \& Melosh 2006).  Work by Snodgrass et al. (2006), while not revealing a definitive rotation period, suggests a value several times longer.  Tidal breakup is implausible given the position of 17P/Holmes (far from any planet or the Sun) at the time of outburst.  A possible trigger for the outburst is a decrease in the perihelion distance from 2.16 AU to 2.05 AU caused by a close approach to Jupiter in January 2004, resulting in an increase in solar insolation (but only by $\sim$ 10\%) to greater depths in the comet's interior.  However, the detailed mechanism by which an increase in insolation might lead to the observed outburst remains unknown.

%can you explain the 5 month delay from perihelion to the time of outburst? ... time for heat to build up.

In this paper we present a set of coordinated observations taken at the 
%University of Hawaii 2.2m telescope and 
Canada-France-Hawaii Telescope (CFHT) in a program designed to monitor the development of the coma in outburst.  A major result is the discovery of multiple sub-nuclei ejected from Comet 17P/Holmes during the October 2007 outburst.  We discuss their dynamical and physical characteristics and the constraints placed by their existence on the outburst mechanism.

\section{Observations and Data Reduction}

%\subsection{Data from the University of Hawaii 2.2m Telescope}

%We used observations taken on UT 2007 Oct.\ 26 with the University of Hawaii 2.2m telescope, also on Mauna Kea.  A Tektronix 2048$\times$2048 pixel CCD with a pixel scale of $0^{\prime\prime}.22$ pixel$^{-1}$ was used to image the nucleus in R band with exposure times of only 0.5 s to avoid saturating the nucleus.  The telescope was auto-guided on a nearby star and tracked at non-sidereal rates.  These short exposures contained a signature from the shutter, which was removed using flatfields with similar exposure times.  We median-combined five of these frames to produce an image devoid of cosmic-rays.  The magnitude of the nucleus was calibrated using observations of Landolt stars (\cite{1992AJ....104..340L}).

%\subsection{Data from the Canada-France-Hawaii Telescope}

We obtained images in the SDSS r' filter ($\lambda_{c}$ = 6250 \AA) on UT 2007 Nov.\ 6, 8-15 at the 3.6 m CFHT atop Mauna Kea.  The instrument used was MegaCam, a wide-field mosaic camera of 36 CCDs that covers a square-degree field of view (Boulade et al. 2003).  Each chip in MegaCam has 2048 $\times$ 4068 pixels, with an image scale of $0^{\prime\prime}.185$ pixel$^{-1}$.  Two sets of five images were taken on each night using a standard dithering pattern to cover the $80^{\prime\prime}$ gaps between chips.  The first set had individual exposure times of 50 s and was intended to provide deep imaging of the comet, while the second set had exposure times of 5s to provide unsaturated photometry near the nucleus.  Time was allocated through a target-of-opportunity program and images were obtained in queue-scheduled mode.  Although non-sidereal tracking was unavailable, trailing losses are not significant in our data as the comet traversed only $0^{\prime\prime}.4$ during the 50 s exposures - less than the full-width half maximum, which was typically $0^{\prime\prime}.85$.  Pre-processing was done by the Elixir pipeline (Magnier \& Cuillandre 2004) which removes the instrumental signature using bias frames and twilight flatfields.  For improved astrometric calibration, we resampled the images using SWarp, released by the Terapix data center at the Institut d'Astrophysique de Paris, and attained $0^{\prime\prime}.2$ astrometric accuracy.  The dithered images were median-combined using IRAF to produce a contiguous field of view.   The weather was seen to be photometric by the CFHT Skyprobe on every night except UT 2007 Nov.\ 6.  Using field stars that could be found in successive images we performed relative photometry across the nine nights of data and found that all nights were consistent to within $\sim$ 0.1 mag.  To account for varying levels of extinction between nights we normalized all measurements to those on UT 2007 Nov.\ 9.  Fluxes were converted to calibrated magnitudes using instrumental zero-points calculated by the Elixir pipeline using Landolt fields (Landolt 1992).

Table~\ref{table:obs} provides a journal of observations.  A sample image (before spatial filtering) is shown in Figure \ref{fig:beforeafter} (left panel).

%\section{Radially Expelled Subnuclei}

\subsection{Spatial filtering of images}

 The very large dynamic range of the coma hinders detection of small
embedded features.  Therefore, we elected to filter the images to suppress the
coma and bring out small--scale fluctuations.  Various algorithms to do so exist in the literature (Larson \& Slaughter 1992; Schleicher \& Farnham 2004), notably the Larson-Sekanina radial and rotational shift-difference algorithm
(Larson \& Sekanina 1984), which detects asymmetries by
subtracting each image pixel from a neighboring pixel separated by
offsets $\Delta R, \Delta \theta$ in nucleus--centered polar
coordinates. Different choices of offsets produce sensitivities to
various kinds of features.  Such a filter has been used on images of 17P/Holmes by Moreno et al. (2008).
%The Larson-Sekanina method maps a nucleus--centered polar coordinate image
%$I(R,\theta)$ to a new image $2I(R,\theta)-$$I(R-\Delta R,\theta-\Delta\theta)$-
%$I(R-\Delta R,\theta+\Delta\theta)$, suppressing broad features and emphasizing
%narrow ones.   

We convolved the images with a
Laplacian filter (Figure \ref{fig:beforeafter}), consisting of a positive Gaussian nestled inside a
broader negative Gaussian. This type of filter is traditionally used
in image processing as an edge detector, signaling regions where the
surface brightness gradient changes.  Unlike the Larson--Sekanina
filter, this filter is anisotropic and does not assume a center to the
coordinate system. The only free parameters are the radii of the two Gaussians.  The Laplacian
filter is akin to a negative second derivative, and thus produces a
positive signal on locally concave portions of the image, like bright
trails or fragments.  After exploring a range of filter scales, we
found an inner Gaussian one-$\sigma$ radius of 3 binned pixels
and an outer radius of 6 binned pixels provided greatest sensitivity to the embedded features.  A binned pixel is
$1^{\prime\prime}.85$.  Figure \ref{fig:multilumpimg} shows the result of applying the Laplacian filter to our images.

For comparison, we also applied the Larson-Sekanina method to our images, with $\Delta R=5$ binned pixels ($9^{\prime\prime}.25$), and
$\Delta \theta=5^\circ$.  Figure \ref{fig:sekanina} shows the resulting images.  A box-car median has been subtracted from the images to improve contrast.

The coma largely vanished in our images when convolved with the Larson-Sekanina algorithm and the Laplacian filter, leaving behind a network of apparent dust trails and
possible fragments.  While most fragments and dust trails identified in these images can also be identified in those images processed with the Larson-Sekanina filter we find that background objects are better suppressed by the Laplacian filter.  To minimize false-detections, we choose to use the Laplace filter on our widefield images of 17P/Holmes.

%In this work, however, we elect to convolve the images with a
%Laplacian filter, consisting of a positive Gaussian nestled inside a
%broader negative Gaussian. This type of filter is traditionally used
%in image processing as an edge detector, signaling regions where the
%surface brightness gradient changes.  Unlike the Larson--Sekanina
%filter, this filter is anisotropic and does not assume a center to the
%coordinate system. The only free parameters are the radii of the two Gaussians.  The Laplacian
%filter is akin to a negative second derivative, and thus produces a
%positive signal on locally concave portions of the image, like bright
%trails or fragments.

\subsection{Detection of fragments}

We wrote software to cross--identify persistent brightness maxima in
our images.  This software displays a Laplacian-filtered
image, allows a user to select an apparent fragment, and then scans a
$7\times7$ binned pixel region for the brightest $3\times3$ box to find the
putative fragment center. We linked candidates from
image to image by hand-selecting the same bright regions found in the
previous image.  Hence, any motion observed should be genuine, because
we allow the peak finding algorithm to hunt for the fragment center.  In all, we found 16 fragment candidates, labeled A--P in Figure
\ref{fig:multilumpimg}.

\section{Characterization of the Fragments}
\subsection{Aperture Photometry}

The fragments were originally identified using images processed with a Laplacian filter to highlight small-scale fluctuations.  To investigate their photometric properties we used our set of unfiltered long-exposure images.    

To minimize background contributions we subtracted the signal from the spherical dust shell surrounding the nucleus.  We computed the median of concentric annuli centered on the nucleus, used cubic spline interpolation to calculate a functional form for the coma and subtracted the resulting fit.  Using the positions of the detected fragments we centered apertures on 89 locations in the 8 nightly images.  The radius of the aperture was varied between $0^{\prime\prime}.185$ and $9^{\prime\prime}.250$, corresponding to physical distances of 145 km and 7265 km, respectively.  The background coma level was calculated using an annulus with an inner radius of $11^{\prime\prime}.1$ and an outer radius of $14^{\prime\prime}.8$.  For the following analyses we choose to use an aperture with a radius of $2^{\prime\prime}.22$.  Larger apertures tend to introduce large errors from imperfect background subtraction, while smaller apertures fail to include much of the light reflected by the fragments.

\subsection{Background Comparison}

First, we performed aperture photometry on a large sample of background regions to determine if the detected fragments were statistically brighter than the background.  These background apertures were centered on a total of 890 points across the 8 images with approximate distances from the nucleus equal to those of the detected fragments and at random position angles.

\begin{figure}
\centering
\includegraphics[scale=0.75]{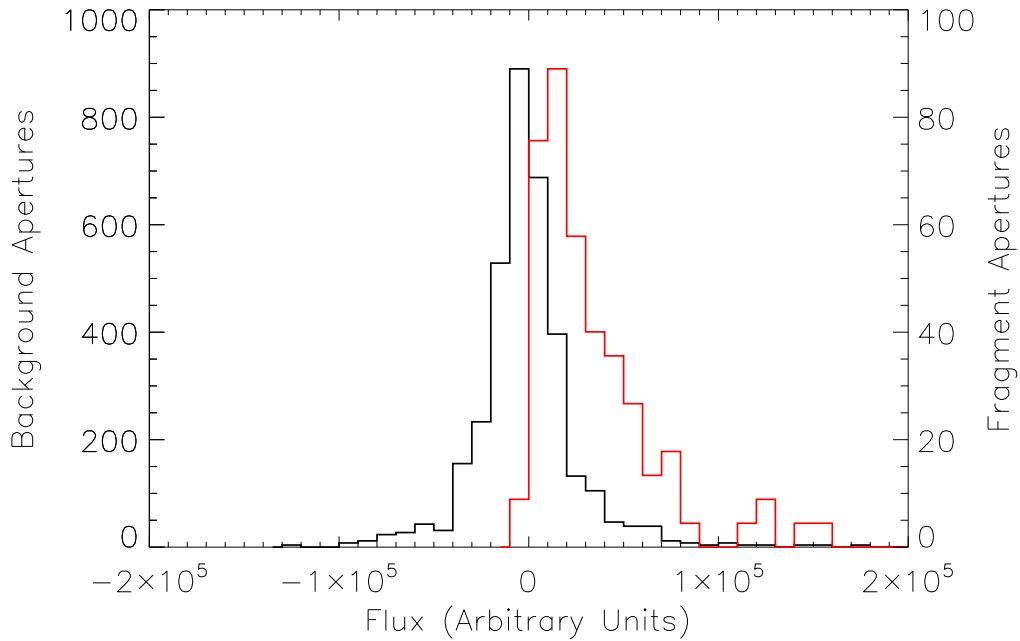}
\caption{Comparison of fluxes contained within 890 $2^{\prime\prime}.22$ radius apertures placed on the background (black line, left y-axis) and 89 $2^{\prime\prime}.22$ radius apertures centered on the fragments detected in the Laplacian-filtered images (red line, right y-axis).  Both samples are normalized for the sake of clarity.  The fragments are systematically brighter than the background.  The two samples shown in the histogram have a probability of being drawn from the same population of $\ll$10$^{-4}$.  Fluxes of some fragments fall slightly below zero due to the uncertainties introduced by sky subtraction.}  
\label{fig:aphist}
\end{figure}

The resulting distribution of flux from the background apertures is significantly different from the apertures centered on the apparent locations of the fragments (Figure \ref{fig:aphist}).  Flux contained within the apertures placed on the background is, on average, zero, as one would expect, given accurate sky subtraction, while those apertures centered on the detected fragments, generally, clearly contain an additional source of light.   A Kolmogorov-Smirnov test shows that the probability that the two distributions in the figure are drawn from the same population is $\ll$ 10$^{-4}$, meaning that the fragments are inconsistent with noise.  We acknowledge that the two distributions overlap slightly.  To minimize the impact of false-detections upon the conclusions of this paper, we focus on the ensemble, rather than individual, properties of the fragments.

\subsection{Spatial Convergence of Fragments}

\begin{figure}
\centering
\includegraphics[totalheight=9cm]{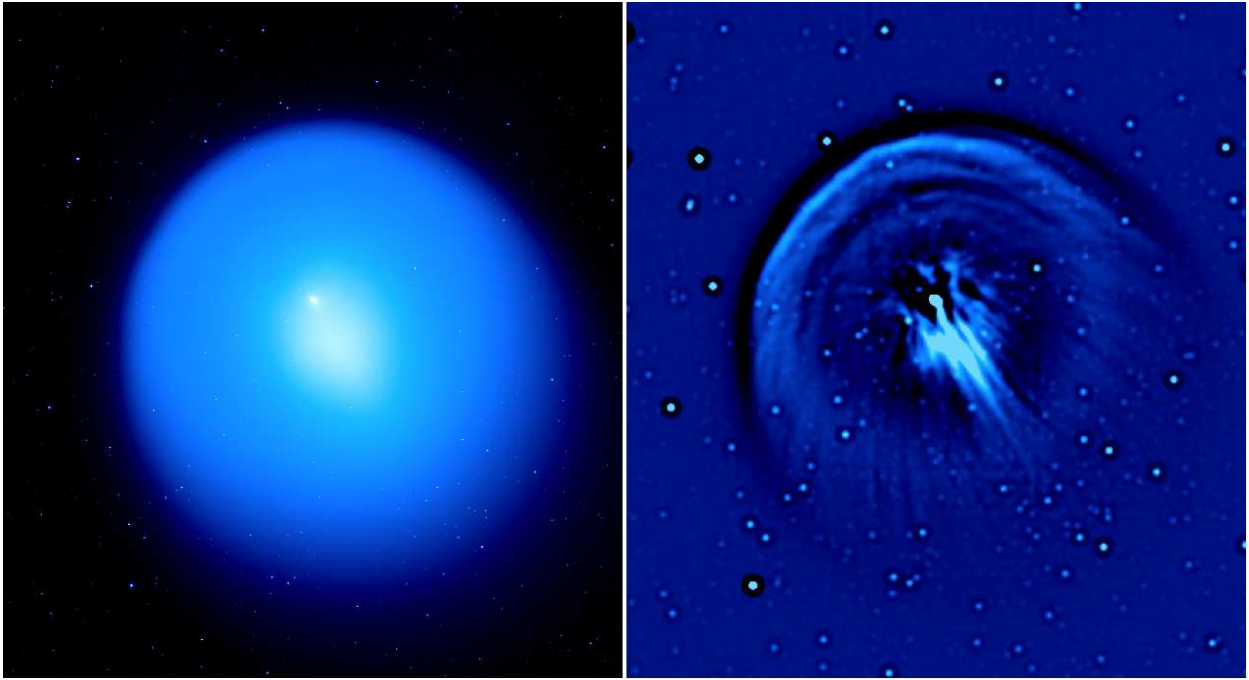}
\caption{Both images are 25.2' $\times$ 27.6', with north up and east to the left.  Left: A 50 second exposure of 17P/Holmes on UT 2007 Nov.\ 6.  The nucleus can be seen north of the center but most morphological features are hidden by the almost-spherical dust shell surrounding it.  Right:  The same 50 second exposure after convolution with a Laplacian filter.  Small-scale features, including dust streaks, background stars, and fragments, are revealed.} 
\label{fig:beforeafter}
\end{figure}

\begin{figure}
\centering
\includegraphics[angle=-90,  totalheight=16cm]{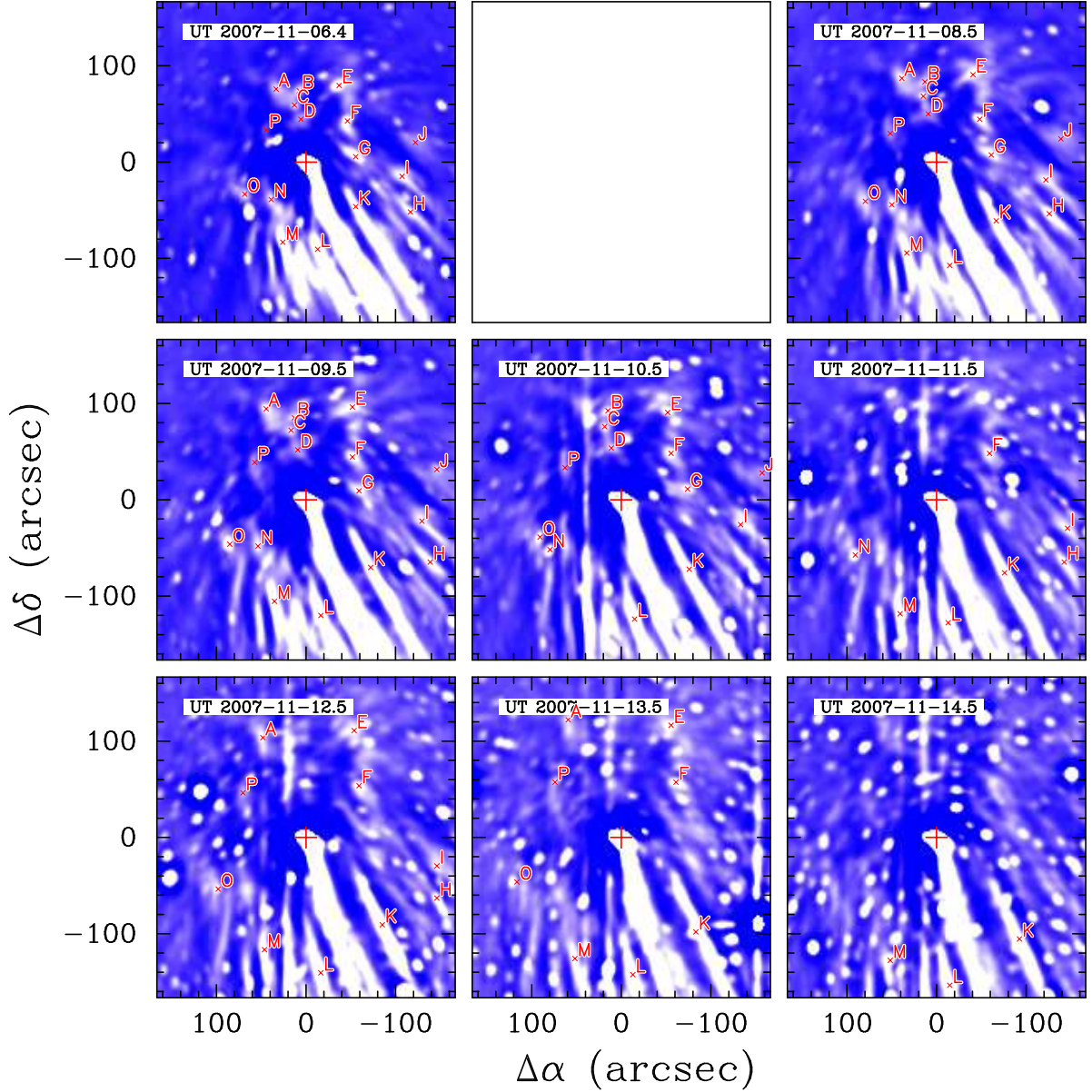}
  \caption{\label{fig:multilumpimg} Laplacian--filtered images of
    Comet Holmes in the usual North up, East left coordinate
    system. There is one image per night, except that no observations
    were obtained on the second night.  The plus symbol indicates the
    fitted position of the nucleus, and the $\times$ symbols, labeled
    A--P, are the fragment positions found through an interactive fitting
    procedure.  Some of the fragments appear to be in the middle of
    extended tails, but with contrast adjustment they do in fact look
    like brighter spots.  Many of the fragments vanish with time, leaving
    only three in the final image above.  Most of the features
    in the image are star residuals, and the vertical streaks are
    remnants of the chip edges from the combining of mosaic images.
    The last night had no detections, and is omitted.}
\end{figure}

\begin{figure}
\centering
\includegraphics[angle=-90, totalheight=16cm]{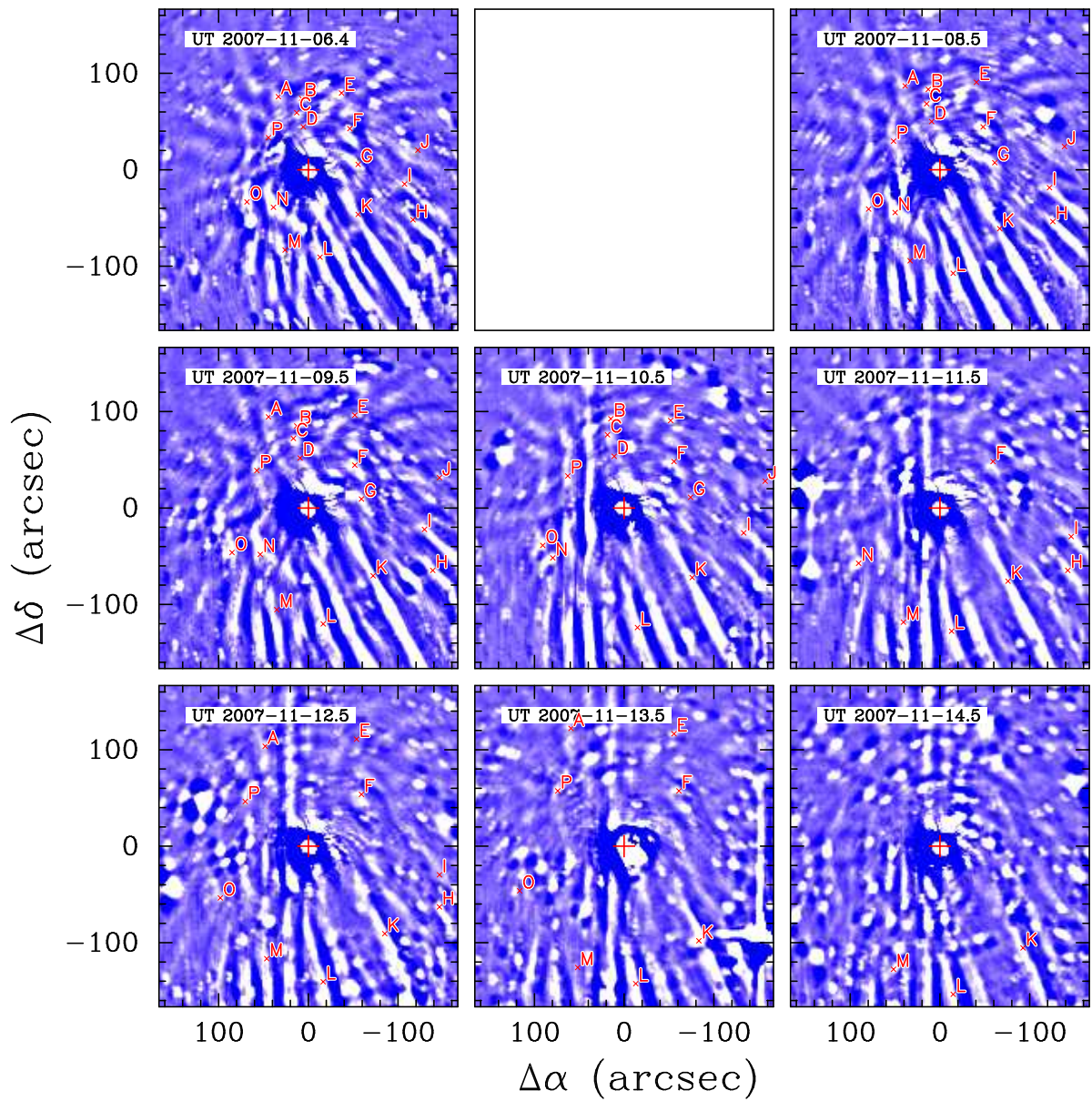}
\caption{\label{fig:sekanina} Images of 17P/Holmes that have been processed with the Larson-Sekanina method in the usual North up, East left coordinate system.  There is one image per night, except that no observations were obtained on the second night.  Background objects, such as stars, are obvious throughout the images, and interfere with attempts to identify potential fragments around the nucleus.  For this reason, we choose to use Laplacian-filtered images, as shown in Figure~\ref{fig:multilumpimg}.}
\end{figure}

The procedure we have described is, of necessity, subjective.  It is
difficult to distinguish between a star residual and a tail-less fragment,
for example, and it is possible that some of our fragments are simply
unconnected random variations in the coma.  However, when Figure
\ref{fig:multilumpimg} is viewed as an animation\footnote{Available for viewing as supplementary online-only material} (Figure \ref{fig:beforeafter}), the expanding nature
of the system of fragments is clear. 

To verify that the fragments are real, and are not subjectively selected
artifacts, we examine the average motion to see whether
our fragments converge as expected.  For each apparent fragment, we compute a
velocity by taking the median of velocities from positions on adjacent
nights (Table \ref{table:megafrag}).  We also compute a median position across the first three
images, giving a snapshot of the entire system at the center time of
these images.  

\begin{figure}
\centering
  \includegraphics[angle=-90,  totalheight=12cm]{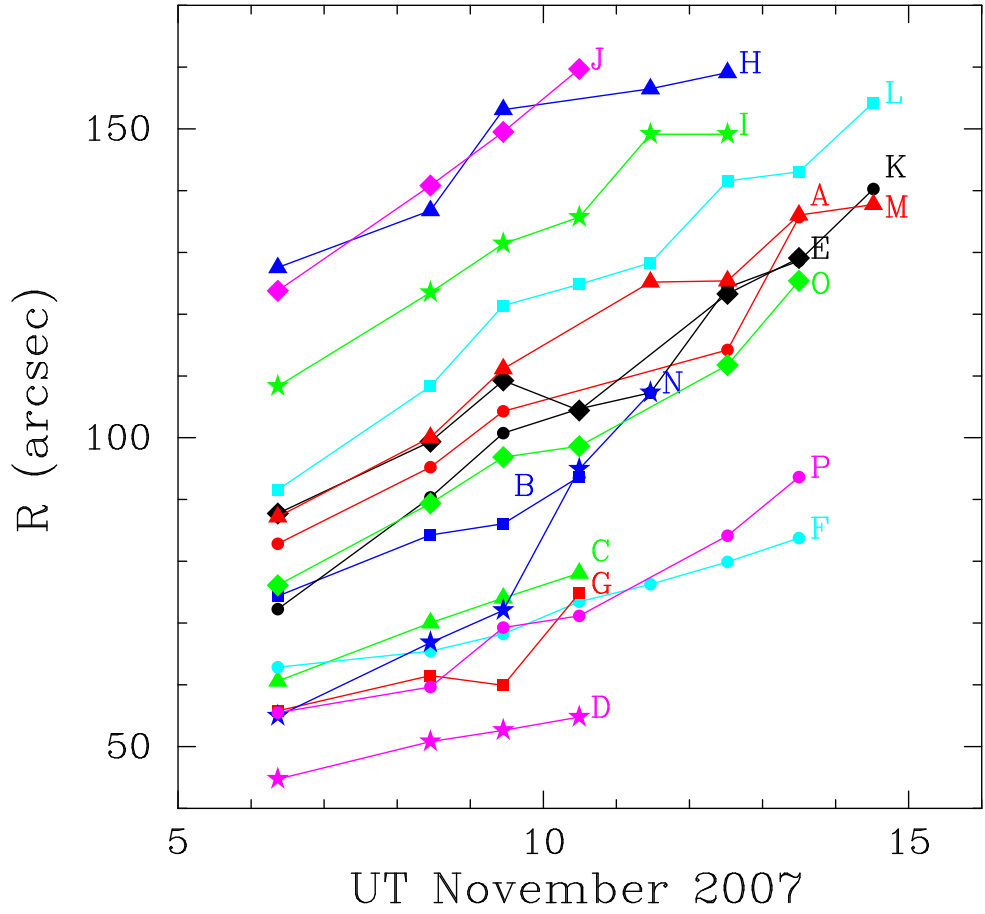}
  %\plotone{lumps.ps}
  \caption{\label{fig:lump-r-vs-mjd} The projected distance $R$
    between each fragment and the nucleus, as a function of time.  }
\end{figure}

\begin{figure}
\centering
  \includegraphics[angle=-90,  totalheight=12cm]{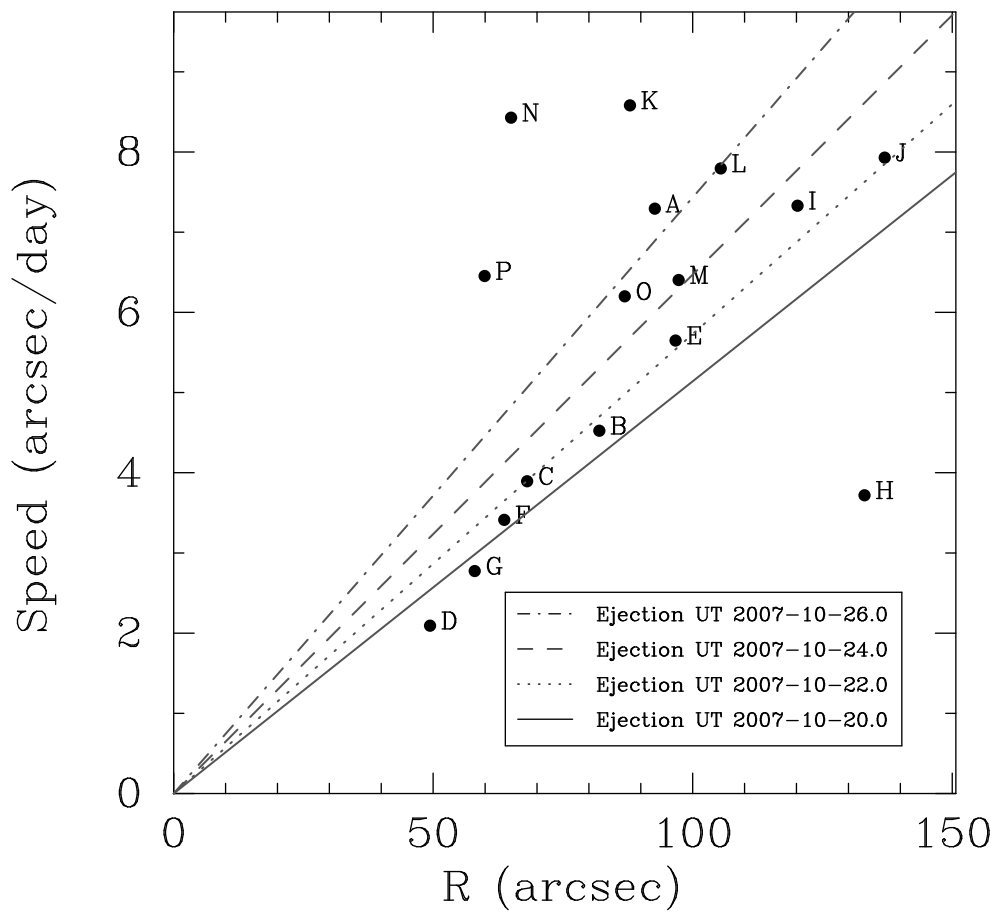}
  %\plotone{lumps.ps}
  \caption{\label{fig:lump-r-vs-vel} The velocity versus radial
    position of the fragments plotted in Figure \ref{fig:multilumpimg}, with the
    lines representing the velocity--radius relationship that causes
    convergence at a particular date.  The relationship between velocity and $r$ is statistically
    significant at the $p=0.06$
    level according to Spearman's $r$ test.}
\end{figure}

We expect any radially expanding system originating at a single time
and place to show a linear relationship between velocity and radius.
Indeed, Figures \ref{fig:lump-r-vs-mjd} and \ref{fig:lump-r-vs-vel}
show such a relationship to be present in our data. There is a
positive correlation between radius and velocity at the Spearman
rank-order probability $p_{\rm SRO}=0.062$.  

We next extrapolate the fragment's velocity back in time from the
fragment's median position at its median observation time. Figure
\ref{fig:lumpxy} shows the individual fragments plotted on a single graph,
with their motion extrapolated back in time.  It is evident that some
of the measurements may be spurious or inaccurate, but on average the
fragments move radially outward in time.

\begin{figure}
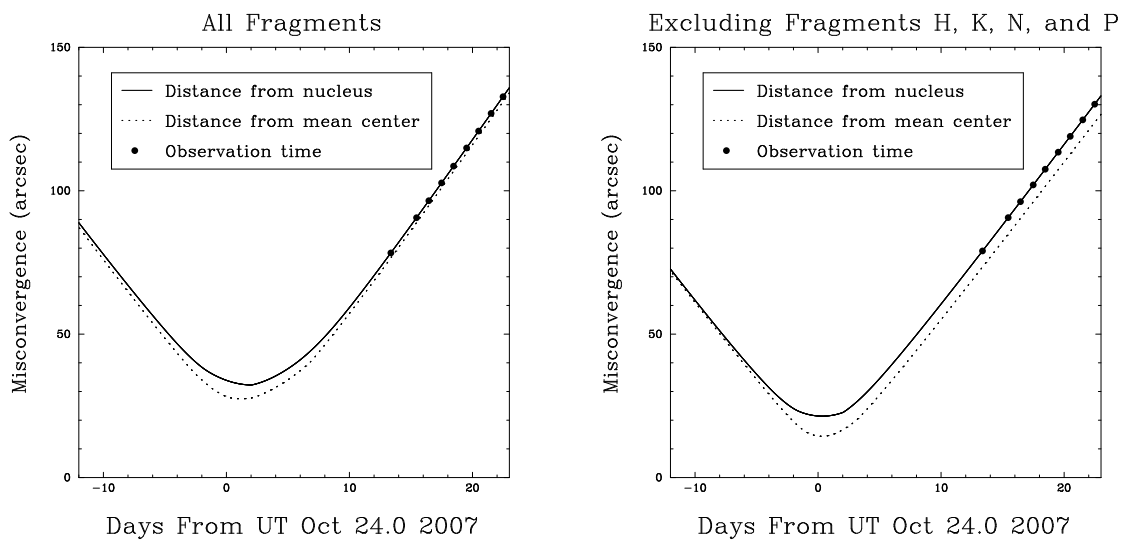

\centering \includegraphics[angle=-90,  totalheight=7cm]{center-miss-alldata.ps} 
\hskip 30pt
\centering \includegraphics[angle=-90,  totalheight=7cm]{center-miss-gooddata.ps} 
  \caption{\label{fig:lumpconverge} The mean distance of the fragments
    from the nucleus and from their mean center, extrapolated
    backward in time. Points indicate days on which we obtained
    observations. The left panel uses all the fragments, and the right
    panel excludes potential outlier fragments $H, K, N$, and $P$ as identified from Figure
    \ref{fig:lump-r-vs-vel}. The times of convergence are in $2\sigma$
    ($0.5\sigma$) agreement with published estimates of the outburst
    date for the complete (truncated) data sets.}
\end{figure}

Finally, we estimate the convergence time of the entire fragment ensemble
using the median positions and velocities.  Table \ref{table:lumpapproach} shows the time and distance of closest approach to the nucleus for each fragment.  Figure
\ref{fig:lumpconverge} shows the median distance of the fragments from
both the nucleus, and from their common center.  For the entire data
set, the fragments converge closest to the nucleus at UT Oct $26.0\pm1.0$ and
closest to their common center at UT Oct $25.2\pm 1.0$, where the
uncertainty is computed through a bootstrapping procedure.  This time
is within $2\sigma$ of the likely eruption time of Oct $23.7\pm
0.2$ (Wilkening et al. 2007).  If we omit fragments
$H, K, N$, and $P$, which fall off the linear relation in Figure
\ref{fig:lump-r-vs-vel}, then the closest convergence time with
respect to both the nucleus and the common center becomes UT Oct $24.3\pm1.2$,
in excellent agreement with Wilkening et al. (2007) and
Hsieh et al. (2007), who obtained UT Oct $23.7\pm0.2$ and UT Oct
$23.8$ (no error given), respectively.     We conclude that the
fragments not only radiate outward, but their positions converge at the time of the
initial outburst, lending credence to the hypothesis that we are observing
pieces of debris from the original outburst.

\section{Discussion}
\subsection{\label{subsec:phasespace} Phase Space Distribution of Fragments}

\begin{figure}
\centering
  \includegraphics[angle=-90,  totalheight=12cm]{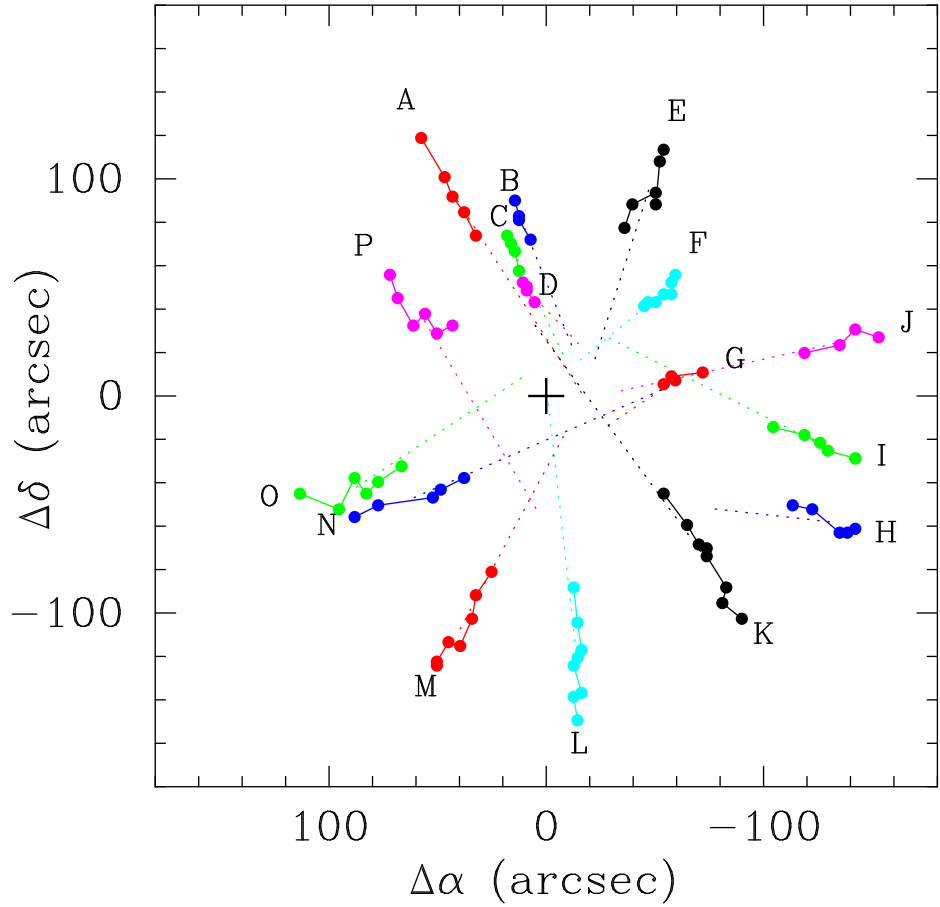}
  %\plotone{lumps.ps}
  \caption{\label{fig:lumpxy} The positions of the fragments in Figure
    \ref{fig:multilumpimg}, with dotted lines representing the
    extrapolation of each fragment back in time, based on the fragment's
    median velocity and position. Although the individual positions
    are noisy, the ensemble of fragments converges close to the
    nucleus.}
\end{figure}

The position angle of the fragments measured from the nucleus appears
uniformly distributed (Figure \ref{fig:lumpxy}).  This suggests that
the true three--dimensional distribution is either spherical, or a
cone with its axis along the line of sight.
%The latter appears less probable, but remains a possibility because
%the line of sight direction is closely aligned with the anti--solar
%direction.

In a radially expanding system, we expect the true and projected
positional radii and radial velocities to be perfectly correlated.
Comparing both to the same theoretical projected distribution thereby
provides an independent validation of the data.  An important caveat
is that such tests are sensitive to the completeness of the
sample. For example, if the manual peak-finding procedure missed slow
moving fragments near the nucleus, we may understate the level of
central concentration. 

In Table \ref{table:dismodels}, we compare the projected distribution
of the radii and velocities of the fragments with various three
dimensional models. We consider 1) a spherical distribution of
fragments; 2) a model in which fragments lie on an infinitely thin
shell; 3,4) finitely thick hollow shells (or hollow spheres) of
fragments that have 20\% and 50\% of the thickness of the shell's
radius; 5,6) radially symmetric space-filling distributions of
fragments with $r^{-1}$ and $r^{-2}$ number density profiles; 7) a
line of sight hollow cone of fragments; and 8) a line of sight solid
cone of fragments.

For the positions, we use the median radius and
time of the first three nights, and for the velocities, we use only
the radial component of the best--estimate median velocity, under the
assumption that any transverse component is noise.  We consider a set
of model distributions consisting of hollow spherical shells of
various thicknesses, filled spheres, and filled and hollow cones.  We
use a Kolmogorov-Smirnov (KS) test to compare our projected $R$ and
$v$ values with the distribution predicted by each model.  This is a
slight misuse of the KS test, because we fix our outermost point to be
at a cumulative probability of 1.  However, our interest is in ruling
out models, and this effect will tend to make all models agree better
with our data.

Table \ref{table:dismodels} shows the KS agreement of the distribution
of the data with the models.  We can rule out the thin shell and the
20\% shell (in which the fragments occupy a shell of thickness equal
to 20\% of the radius) on the basis of both the radial data and the
velocity data.  The 50\% thick shell, solid sphere, and, to a lesser extent,
$r^{-1}$ models are compatible with the data.  If the eruption is
conical rather than spherical, then an edge-enhanced cone is preferred
over one that is solid.  In cases in which the radial data disagree
with the velocity data, like the 20\% spherical shell and the solid
cone, we are inclined to believe that the radial data are more robust.

These results must be interpreted taking into account the completeness
caveat mentioned above. If we assume that fragments closer than
$0.3\times R_{\rm max}$, where $R_{\rm max}$ is the projected radius
of the largest fragment, are invisible to us, then the shell-like
distributions become less likely, but the centrally concentrated
$r^{-2}$ density model and the solid cone are no longer ruled out.  At
most, we can claim that the fragment distribution is not concentrated
at the peripheries, but we cannot rule out a strongly centralized
arrangement.  Our conclusion is that the spatial and radial velocity
distributions of the fragments in the sky-plane are consistent with
isotropic ejection or conical ejection centered around the line of
sight, or very close to it.

It is difficult to envisage a plausible scenario in which 16 fragments
are ejected isotropically from the nucleus, without catastrophic
disruption of the nucleus.  One mechanism has been suggested by
Samarasinha (2001) who proposed that small nuclei ($\sim$~1~km) could
contain connected voids that allow sublimated supervolatiles to move
rapidly through the nucleus.  Assuming these voids have no outlet to
the surface, internal gas pressure could build up until it exceeds the
tensile strength of the mantle.  At this point, an outburst could
occur over a large fraction of the nucleus' surface.  A difficulty
with this model is the very small tensile strength of the cometary
nucleus, which will prevent the build-up of high internal pressures.

% == talk about cone versus isotropy.  why cone might be better.  isotropy suggested by samarasinha but results in catastrophic disruption of the nucleus - not observed here.

%% {\bf NOTE:} Isn't an edge enhanced cone (vs filled cone) the
%% consequence of an impact, which is unlikely, leaving only a
%% filled-spherical model as a likely remaining contender? Also should
%% note that a hemispherical erupution is indistinguishable from a sphere,
%% and represents the meeting point between a cone and an sphere.

%% \subsection{Velocities of lumps}

%% \begin{itemize}

%% \item lumps travel at 100 $m/s$ 5000 Joules/kg - latent heat of
%% fusion is 384,000 Joules/kg, s amorphous to crystalline
%% transformation has the energy (but no mechanism)

%% \item Suppose you have a 10x10 cm 1 kg chunk. To accelerate it to
%%     100 m/s you meed momentum $dP=100 \rm kg m/s$.  A one second
%%     force of $100 \rm kg/m/s^2$ would do it.  This force could be
%%     generated by a pressure of $10^4$ pascal over 0.01 square
%%     meter.  Such a pressure corresponds to 0.01 atomospheres.

%% \end{itemize}

%photometry
\subsection{\label{subsec:sizeestimates} Size Estimates}

At the earliest detection in our data set, the fragments have magnitudes ranging from r' = 17.6 to 23.5 (Table \ref{table:megafrag}).  We consider two models to estimate the sizes of the fragments from the magnitudes.  The apparent magnitude of a monolithic body is related to the viewing geometry and the body's physical characteristics according to the relation

\begin{equation}
g_{\lambda} \Phi_{\alpha} C = 2.25\times10^{22} R^{2} \Delta^{2} \pi 10^{0.4(m_{\sun}-m_{\lambda})} 
\end{equation}

\noindent where $g_{\lambda}$ is the geometric albedo, $\Phi_{\alpha}$ is a function to account for the variation of brightness of the body with phase angle, $C$ [m$^{2}$] is the geometric cross-section of the body, $R$ [AU] and $\Delta$ [AU] are the heliocentric and geocentric distances, and $m_{\sun}$ and $m_{\lambda}$ are the apparent magnitudes of the Sun and the body respectively (Jewitt 1991).  We use a linear approximation for the phase function and set $\Phi_{\alpha}$ = 10$^{-0.4\alpha\beta}$, where $\alpha$ [deg] is the phase angle and $\beta$ [mag deg$^{-1}$] is the phase-coefficient.  We assume a value of 0.035 mag deg$^{-1}$ for the phase-coefficient (Lamy et al. 2004) and $m_{\sun}$ = -26.95 mag when using the SDSS r' filter (Ivezi\'{c} et al. 2001).

%magnitude taken from SDSS calibration (see Ivezic et al. 2001).  obtained from Vsun=-26.75, (B-V)sun=0.65 (Allen 1973) w/ aid of photometric transform from Fukugita et al. 1996.

Model A:  If the detected fragments are monolithic, spherical bodies with geometric albedos of 0.1, we infer that the median radius of a fragment is 1.79 km, with sizes ranging from 0.8 km to 3.0 km (Table~\ref{table:megafrag}).  Since the radius of the parent nucleus is only $\sim$ 1.7 km (Lamy et al. 2000;, Snodgrass et al. 2006) this interpretation can be rejected.  Increasing the albedo to 0.15 yields a range of radii from 0.6 km m to 2.4 km with the median radius being $\sim$ 1.5 km.  We conclude that it is unlikely that the fragments are bare nuclei, and instead proceed to consider the possibility that they are actively-outgassing sub-nuclei.

Model B:  Using the complementary set of 5 second exposures obtained on the same nights at CFHT, we measured the brightness of the unsaturated nucleus, without coma-subtraction.  We find that the apparent magnitude corresponds to an effective radius of $\sim$ 330 km, demonstrating that the dust coma around the nucleus dominates the scattering cross-section, as it appears to do for the fragments we have discovered.  Thus, the magnitude of the fragment ($m_{f}$) or nucleus ($m_{n}$) depends primarily on the amount of dust present in the aperture.  Assuming that the nucleus and the fragments have material sublimating from active regions that cover similar fractions of their surfaces, then the difference of the observed magnitudes is proportional to the ratio of their surface areas, or their radii squared if we assume spherical bodies:

\begin{equation}
10^{-0.4(m_{\rm f}-m_{\rm n})} = \frac{R^{2}_{\rm f}}{R^{2}_{\rm n}}
\end{equation}   

where $R_{f}$ and $R_{n}$ are the radii of the fragment and the nucleus, respectively.  Using this scaling argument, and apparent magnitudes of the fragments and nucleus determined using $2^{\prime\prime}.22$ circular apertures as described in Section 3.1,  we obtain fragment radii between 10 m and 110 m on the first night of detection (Table~\ref{table:megafrag}).  Cometary nuclei typically have active regions that cover only a small percentage of the surface (A'Hearn et al. 1995; Jewitt 2004), due to the gradual formation of an inert mantle by irradiation, micro-meteorite bombardment and loss of volatiles.  It is possible that the fragments were more active than the nucleus.  They may have been rotating rapidly and exposing much of their surface to sunlight, or composed mainly of volatile material with little or no mantle.  Thus, the sizes derived here should be considered as upper limits.

With a density of 400 kg m$^{-3}$ (Richardson \& Melosh) and the scaled radii listed in Table \ref{table:megafrag} for Model B we find that the sixteen fragments have a combined mass of 10$^{10}$ kg, corresponding to $\sim$ 0.1 $\%$ of the mass of a 1.7 km radius, spherical nucleus.  Again, this is an upper limit to the mass in the fragments, and shows that the outburst of 17P/Holmes ejected only a tiny fraction of the total nucleus mass.  
%\citep{2008ICQ....30....3S}.  

\subsection{Acceleration}

We do not detect any systematic acceleration of the fragments between 2007 Nov. 6 and 2007 Nov. 14 UT, since a single mean velocity over the observational dataset predicts a time of ejection that agrees with the published eruption time (Wilkening et al. 2007). This suggests two things: firstly, that radiation pressure does not significantly affect the motion of the fragments and, secondly, that the fragments are not self-propelled by directional sublimation of volatiles.  The first point suggests that the fragments are macroscopic, as opposed to clusters of micron-sized particles, or smaller, that would be easily accelerated in the anti-solar direction by radiation pressure.  The second constrains the nature of the fragments.  Given that these fragments are volatile-rich and actively outgassing (as demonstrated in \S\ref{subsec:sizeestimates}), one may expect self-propulsion to accelerate the fragments in the anti-solar direction, as volatiles would be typically expelled in the sunward-direction.  However, the fragments are likely to be spinning rapidly and may be isothermal,  resulting in sublimation in all directions, and hence no net acceleration.  Thus, the fragments are observed to continue in the directions in which they were ejected, with no noticeable increase in velocity.

\subsection{Correlation of Flux with Radius}

If the observed surface brightness maxima originate from discrete solid fragments
expelled by gas pressure, we expect lighter fragments to be launched
faster, and so to appear at larger radii. Specifically, if we make the
assumption that the observed fragments are icy fragments of size
$\ell_{\rm f}$, density $\rho_{\rm f}$ and mass $m_{\rm f}=\rho_{\rm
  f} \ell_{\rm f}^3$, and are ejected by gas pressure $P$ acting over
a fixed acceleration distance $d$, then the energy transmitted to each
fragment is ${1\over2} m_{\rm f} v_{\rm f}^2= P \ell_{\rm f}^2 d$, and each
fragment's distance to the nucleus $r_{\rm f}$ is given by  $r_{\rm f} \propto v_{\rm
  f} \propto \ell_{\rm f}^{-1/2}$, where $v_{\rm f}$ is the fragment's velocity.  Assuming that a
fragment's brightness is given by its (sublimating) surface area
$\ell_{\rm f}^2$, the expected relationship of $r_{\rm f}$ to photon
flux $f_{\rm f}$ is then $r_{\rm f}\propto f_{\rm f}^{-1/4}$.  

Alternatively, if we change our assumptions so that the gas pressure acts for a fixed
time $t$ instead of a fixed distance $d$, then the imparted momentum
is $P \ell_{\rm f}^2 t = m_{\rm f} v_{\rm f}$, and $r_{\rm f}\propto
f_{\rm f}^{-1/2}$.  In both instances, fragment brightness should be weakly
anti-correlated with non--projected radius $r$.  In a two dimensional projection onto the sky radius $R$, the above  inverse relation will be somewhat washed out.  Nevertheless, we
still expect to find brighter fragments at smaller radii.

\begin{figure}
  \centering 
  \includegraphics[angle=-90,  totalheight=9cm]{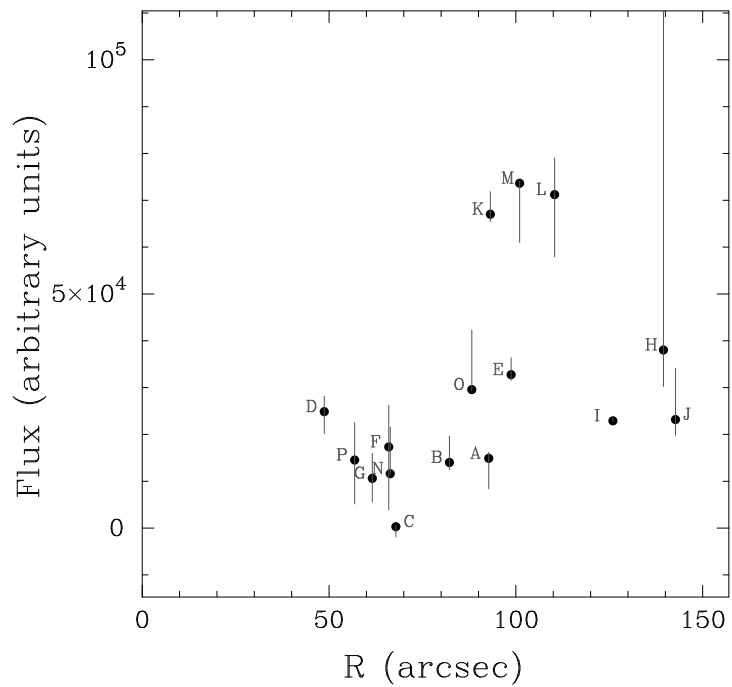} 
  \caption{\label{fig:lump-r-vs-flux} Relationship of the projected
    radial distance from the nucleus, $R$, to each fragment's median
    photometric flux, both quantities averaged over the first three
    nights after accounting for fading. The vertical bars span the
    minimum and maximum of the three nightly fluxes for each fragment.
    There is a statistically significant positive Spearman correlation
    ($p=0.017$) between $R$ and flux, the opposite of what one would
    expect if smaller fragments were expelled at a higher velocity.
  }
\end{figure}

Figure \ref{fig:lump-r-vs-flux} shows the observed relationship of sky radius, $R$,
to flux.  There is a statistically significant (Spearman rank-order
$p_{\rm SRO}=0.017$) positive correlation between $R$ and flux, in
contrast to the expected anti-correlation.  It is reassuring that a
strong deficit of faint fragments at small $R$ was not observed,
because this would be suggestive of a selection bias against finding
fragments in the bright central coma near the nucleus.  The fact that
the shape of the flux distribution varies with $R$ is consistent with
the space-filling distribution suggested by \S\ref{subsec:phasespace},
because a thin expanding shell of fragments would produce a
distribution of fluxes that is invariant in $R$.

In conclusion, the observation that flux increases rather than
decreases with radius argues against a model in which the fragments
consist of monolithic fragments with a sublimation rate and reflective
flux proportional to their surface area.  Otherwise, by simple gas
pressure arguments, one would expect the largest and heaviest
fragments to be closest to the nucleus. Instead, each detected fragment
might in fact be a collection of active objects, rather than a single
cohesive fragment.

%In conclusion, the observation that flux increases rather than
%decreases with radius argues against a model in which the fragments
%consist of single fragments with a sublimation rate and reflective flux
%proportional to their surface area.  We use this observation to suggest that each detected fragment may in fact be a collection of active objects, rather than a single cohesive fragment.

\subsection{Fragment Fading}

\begin{figure}
\centering
\includegraphics[scale=0.90]{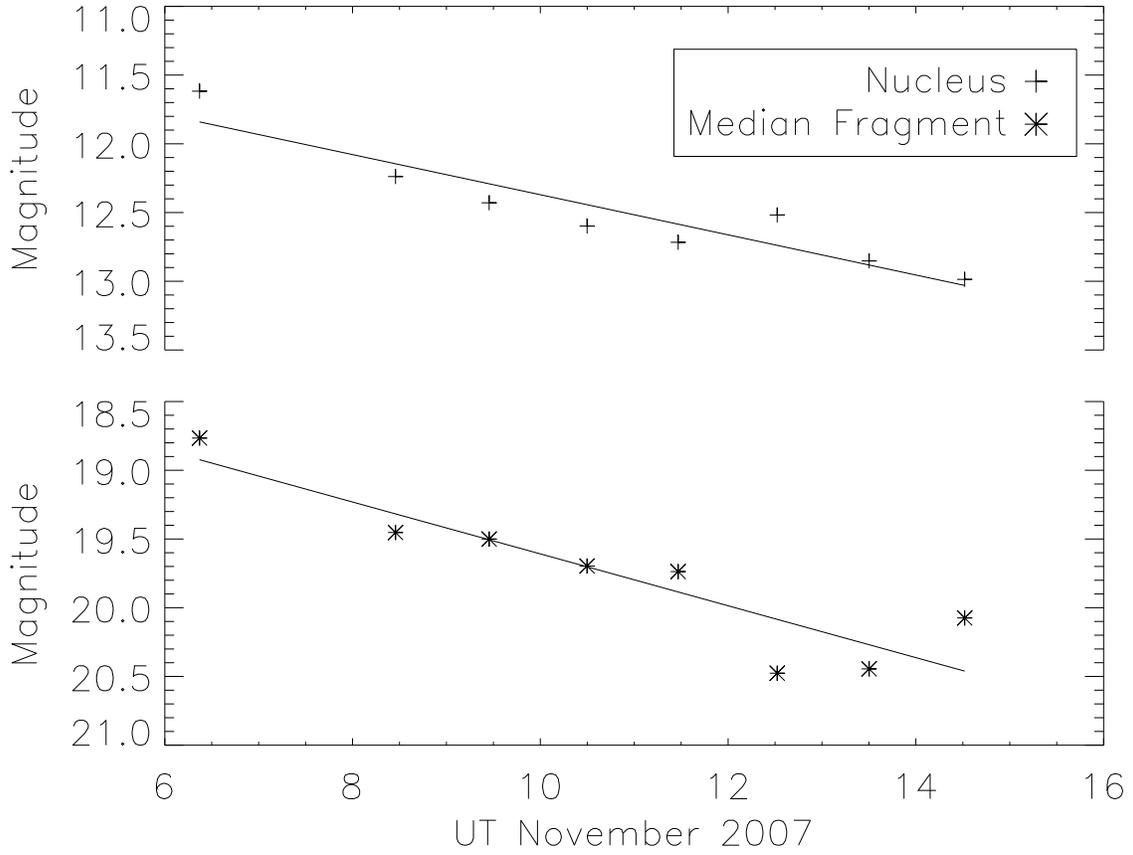}
\caption{Magnitudes of the nucleus and the median fragments as determined from 2.22" aperture photometry.  Linear fits to each line yield fading rates of 0.15 mag day$^{-1}$ and 0.19 mag day$^{-1}$ for the nucleus and average fragment respectively.}
\label{fig:fading}
\end{figure}

\begin{figure}
  \centering 
  \includegraphics[angle=-90,  totalheight=7cm]{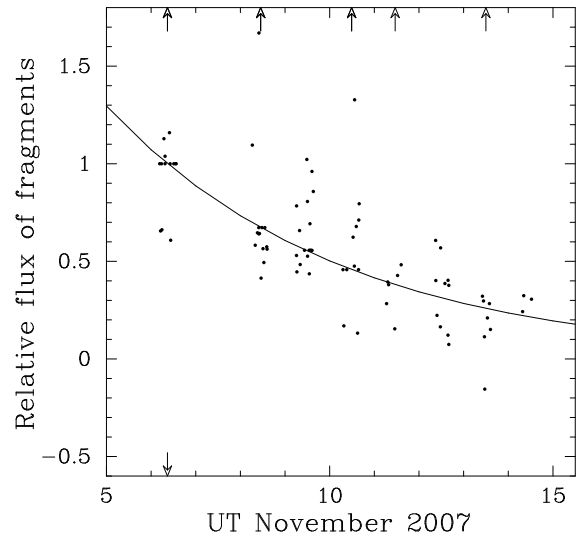} 
  \caption{\label{fig:lump-fading} Fading of the fragments with
    time. Based on the best exponential fit of the entire data set,
    the flux of each fragment at each measured time was rescaled so
    that the fits of all fragments have a flux of 1.0 on the first
    night, and then plotted as a point.  The solid curve is the
    best--fitting exponential $\exp(-0.18 \times {\rm time})$, where time is in days.  Arrows
    indicate outliers falling off the plot.  For the purpose of
    plotting only, the x-axis values have been randomized slightly to
    prevent points from overlapping.}
\end{figure}

Figure~\ref{fig:fading} shows the temporal fading of the nucleus and the median of the fragments' magnitudes during our observations.  The nucleus fades at a rate of 0.15 mag day$^{-1}$ while the fragments, on average, fade at a similar rate of 0.19 mag day$^{-1}$.  Figure \ref{fig:lump-fading} rescales the flux of each fragment to a common baseline
using an exponential fit with a shared time constant.  Over the nine days
plotted, the fragments fade by about 80\%.

We checked the functional form of the fade and find that an exponential fit is best, but that linear and quadratic fits cannot be ruled out.
We evaluate the significance of differences among the linear, quadratic, and exponential models by
bootstrap resampling the data set. We resample from the set of
fragments themselves, to create simulated data containing the same
number of fragments, but with possible repeats of individual
fragments' time series.  In 5000 resamplings, the quadratic model is
favored over the linear model in 99.1\% of instances; the exponential
model is favored over the linear model in 98.6\% of instances; and the
exponential model is favored over the quadratic model in 97.9\% of
instances. We conclude, with $\gtrsim 2\sigma$ certainty, that the
fading of the fragments is best described by an exponential law, or a
constant fractional fading per unit time.  In Figure
\ref{fig:lump-fading} we plotted the best exponential falloff rate of
$0.18\,{\rm day}^{-1}$, representing a 5.6 day exponential time scale.

We solved the sublimation equilibrium equation for icy grains with Bond albedo 0.1 to find maximum sublimation rates of $\sim$ 5 $\times$ 10$^{-8}$ m s$^{-1}$ (for a flat slab perpendicular to sunlight) to $\sim$10$^{-9}$ m s$^{-1}$ (for an isothermal sphere).  Using these sublimation rates and the exponential time scale of 5.6 days, we calculate fragment sizes of 5 $\times$ 10$^{-4}$ m to 0.02 m.  These sizes are much too small to account for the observed magnitudes of the fragments, unless each fragment observed is in fact a collection of icy grains.

In Appendix A, we show that a power law distribution of sub-fragments sublimating
at a constant rate naturally produces an exponential decay in emission, as is observed.

\section{Physical processes of ejection}

% need segue

% discussion of Samarasinha's paper needs to go in here

The relative velocities of the fragments are puzzlingly high.  Typically fragments ejected from short-period comets have separation velocities of a few meters per second (comparable to the nucleus escape velocities; Boehnhardt 2004) but our measurements show typical velocities of $\sim$100 m s$^{-1}$ on the sky-plane.  We note that observations of split comets are usually performed weeks or months after the event, at which time high-speed fragments that are fading, like those considered here, would no longer be detectable.  Thus, the deficit of small, active, high-velocity fragments around other split comets may be due to observational biases, and the nature of the material ejected during the 2007 outburst of 17P/Holmes may not be particularly unusual.  However, the mechanism responsible for accelerating these fragments to velocities of $\sim$ 100 m s$^{-1}$ is difficult to establish.  We consider several possibilities.
 
Could rotational fragmentation account for the measured velocities?   The shortest possible rotation period for 17P/Holmes is 5.2 hours.  If the nucleus rotates faster than this then it will break up, assuming it is a strengthless body.  This rotation period corresponds to a surface velocity of just 0.6 m s$^{-1}$, ruling out rotational disintegration as mechanism of expulsion.

We also consider the possibility that the fragments were expelled slowly, and then accelerated like rockets by the reaction against their own sublimation.   The first problem with this ``rocket'' model is that acceleration would be in the anti-solar direction, meaning that the true velocity must be four times greater than the largest $\sim 100\,\rm m\,s^{-1}$ spatial velocity observed, because the line of sight is nearly parallel to
the motion, and only a small component of the velocity
appears as tangential motion on the sky.
Thus, there is no way to account for  the $\sim 100\,\rm m\,s^{-1}$ transverse velocities observed when the transverse motion
is not aligned with the projection of the anti-solar direction. 

Moreover, it is difficult to account for the absolute speed of the fragments using rocket propulsion. The rocket equation gives, for a rocket of initial mass $m_i$, final mass $m_f$, and  exhaust velocity $v_e$, a rocket velocity $v_r = v_e \ln (m_i/m_f)$.  For a non-rotating sublimating fragment, however, the exhaust is emitted over a hemisphere, rather than through a rocket nozzle, so that half the momentum is lost, and $v_r =  v_e \ln (m_i/m_f)/2 $.  Assuming that the exhaust velocity is given by the $(3/2) k_B T$ energy of water molecule at a $190\,\rm K$ sublimation temperature, and allowing no loss of energy into the water molecules' rotational modes, $v_e=511\,\rm m\,s^{-1}$.   To achieve a final velocity of $400 \,\rm m\,s^{-1}$, 84\% of the fragments' mass must be  sublimated as rocket fuel, implying that their radius decreases by half during acceleration.  At the fastest plausible sublimation rate of $5\times10^{-8}\,\rm m\,s^{-1}$, taken to occur over a ten day acceleration period, this fraction implies a maximum initial fragment radius of 0.08 m.   However, the thermalization time of a fragment of ice of this size is about half a day, so that the fragments would quickly become isothermal and the asymmetrical sublimation that provides the propulsion would cease long before the necessary velocity was reached.  

So, the measured ejection velocities are difficult to explain. 
The most plausible ejection method is through gas pressure.  In the simplest possible model,
a pressure $P$ ejects a mass of linear size $L$, acting over a distance of $L$ before the 
gas dissipates.  Such a process would resemble an explosion occurring under the fragments.
If one assumes porous fragments with a density $\rho= 400\,\rm kg\,m^{-3}$,  the pressure required
is given by equating work done, $P L^2 L$, with kinetic energy, $\rho L^3 v^2 /2$, giving
$P=\rho v^2 /2 \sim2\times10^6\,{\rm Pa}$.    It is possible for CO gas to create such a pressure.

It has been hypothesized that  a runaway crystallization of amorphous ice may provide enough energy to cause CO ice to sublimate, but this
scenario presents many problems in the context of 17P/Holmes. If this is a surface explosion, then the explosive reaction must 
propagate through the medium at approximately the fragment velocity of $100\,\rm m\,s^{-1}$, in 
order for the gas to push the fragments before it dissipates.   However, the transition heats the ice only by less
than $40\,\rm K$.  For the transition to propagate at explosive speeds, the reaction time $\tau$ must be under a millisecond,
given a propagation speed of $\sqrt{\kappa/\tau}$, where $\kappa\sim$ 3 $\times$ 10$^{-7}$ m $^{2}$\,s$^{-1}$ is an upper bound for the
thermal diffusivity of ice.  However, it is difficult to heat the ice from a highly stable state where $\tau$ is days or weeks, to a state in which it is so unstable that $\tau<3\times10^{-11}$, using only the energy of the transition.

It may be possible to mitigate some of these difficulties by postulating that the acceleration distance is much
larger than the fragment size.  For example, there may be a broad gas emitting region of size $h$ on 17P/Holmes, creating a 
flux of gas normal to the surface, so that fragments would feel a push 
for a larger  distance $\sim h$ from the surface.  The pressures required would then be reduced to 
$P \sim \rho (L/h) v^2 /2$.    This might reduce the amount of gas flux required, but may not remove the need
for a fast propagating reaction. 

Alternatively, we could suppose that fragments are accelerated along an extended trajectory, as in a vent.
In this case, we would expect the following equation of energy conservation to hold, neglecting friction and gravity:

\begin{equation}
{1\over 2} v^2 = \int_{P_i}^{P_i} {dP\over\rho}
\end{equation}
If the fluid is a mixture of gas and solid, with gas fraction $f$ by weight, then the overall density is
\begin{equation}
\rho=\left(  {f\over\rho_{\rm gas}} + {1-f  \over \rho_{\rm solid}}      \right)^{-1}
\end{equation}
Assuming that the solids dominate, and act as a thermal reservoir that keeps the expansion isothermal at temperature $T$, 
$P= \rho_{\rm gas} R_g T$, where $R_g$ is the universal gas constant.  Then the final fluid velocity is given by
\begin{equation}
{1\over 2} v^2 = {1-f  \over \rho_{\rm solid}}  (P_i - P_f) + f R_g T \ln (P_i/P_f)
\end{equation}
If one assumes that the pressure arises from a conversion of mass fraction $f=0.1$ CO, that $\rho_{\rm solid}$= 400 kg m$^{-3}$, and that the pressure falls to 1/3 of its initial value when the fluid reaches the surface, then the final velocity is $70\,\rm m\,s^{-1}$ for $f=0.10$, and $23\,\rm m\,s^{-1}$ for $f=0.01$.   Changing the pressure at the surface affects the final velocity
only modestly.  It appears that a 10\% CO fraction could provide just enough gas production for the velocities observed,
assuming eruption from a vent.

\section{Summary}

We have identified and characterized fragments that were ejected from the nucleus of 17P/Holmes during its spectacular outburst in October 2007.  Our findings are as follows:

% don't underestimate importance of simple existence of fragments. 

\begin{enumerate}

\item Sixteen fragments are detected in Laplacian-filtered images where the coma has been suppressed using an azimuthal average.

\item The motion of the fragments implies either isotropic or conical ejection from the nucleus on UT 2007 Oct.\ 24.3 $\pm$ 1.2.

\item Results from aperture photometry are inconsistent with inert, monolithic bodies.  Modeling the fragments as sublimating cometesimals yields radii of 10 m to 110 m.  Assuming a density of 400 kg m$^{-3}$, the fragments account for 10$^{10}$ kg of the total mass ejected, or $\sim$ 0.1 \% of the nucleus mass.

\item The fragments move unusually fast, with on-sky velocities of up to 125 m s$^{-1}$.  Acceleration by CO (or other supervolatile) gas drag forces might be able to generate such large velocities given appropriate launch conditions at the nucleus.

\item We detect no systematic acceleration of the fragments and deduce that the bodies are not self-propelled by sublimation in a preferred direction.

\item The fragments fade at a rate of $\sim$ 0.19 mag day$^{-1}$, consistent with the idea that they are active bodies, eventually becoming inert as surface volatiles are depleted.   

% lifetimes too (at least, when do they disappear), but make sure this is mentioned in discussion

\end{enumerate}

The authors thank CFHT Director Christian Veillet for allocating time to this target of opportunity program and Pierre Martin, Jean-Charles Cuillandre, and the QSO team at CFHT for providing observational assistance.  Pedro Lacerda and Bin Yang provided helpful comments.  We appreciate support from a NASA Outer Planets Research grant to DJ.

%have presented detection of 16 fragments moving radially outward from the Holmes parent nucleus one week after initial outburst
%isotropic ejection
%size/magnitudes/really detected!
%motion consistent with having originated from the nucleus at time of outburst
%observed fading

\newpage

\section{References}

%\bibliographystyle{aj}
%\bibliography{masterbib}

%\bibitem[data]{key}

%\bibliography{/Users/rach/masterbib}

% references by hand ....
A'Hearn, M. F., Millis, R. L., Schleicher, D. G., Osip, D. J., Birch, P. V. 1995, Icarus, 118, 223

Boehnhardt, H. 2004, in \textit{Comets II}, ed. M. C. Festou et al. (Tucson: Univ. Arizona Press), 301

Boulade, O., et al. 2003, Proc. SPIE, 4841, 72 

Buzzi, L., Muler, G., Kidger, M., Henriquez Santana, J. A., Naves, R., Campas, M., Kugel, F., Rinner, C. 2007, IAU Circ. 8886

Dohnanyi, J. S. 1969, J. Geophys. Res., 74, 2531

Hsieh, H. H., Fitzsimmons, A., Pollacco, D. L. 2007, IAU Circ. 8897

Holmes, E. 1892, The Observatory, 15, 441

Ivezi\'{c}, Z. et al. 2001, AJ, 122, 2749 

Jewitt, D. C. 1991, in \textit{Comets in the Post-Halley Era}, 1, 19

Jewitt, D. C. 2004, in \textit{Comets II}, ed. M. C. Festou et al. (Tucson: Univ. Arizona Press), 659

Lamy, P. L. Toth, I., Fern\'{a}ndez, Y. R., Weaver, H. A. 2004, in \textit{Comets II}, ed. M. C. Festou et al.  (Tucson: Univ. Arizona Press), 223

Lamy, P. L., Toth, I., Weaver, H. A., Delahodde, C., Jorda, L., A'Hearn, M. F. 2000, BAAS, 32, 1061

Landolt, A.U. 1992, AJ, 104, 340

Larson, S.M., Sekanina, Z. 1984, AJ, 89, 571

Larson, S. M., Slaughter, C.D. 1992, Proc. ACM, 337 

Magnier, E. A., Cuillandre, J.-C. 2004, PASP, 116, 449

Moreno, F., Ortiz, J. L., Santos-Sanz, P., Morales, N., Vidal-N\'{u}{\~n}ez, M. J., Lara, L. M., Guti\'{e}rrez, P. J. 2008, ApJL, 677, L63

Palisa, F. 1893, Astron. Nachr, 132, 31

Richardson, J. E., Melosh, H. J. 2006, Abstract 1836, Lunar and Planetary Science Meeting XXXVII

Samarasinha, N. H. 2001, Icarus, 154, 540

Schleicher, D.G. 2009, AJ, 138, 1062

Schleicher, D.G., Farnham, T.L. 2004, in \textit{Comets II}, ed. M. C. Festou et al.  (Tucson: Univ. Arizona Press), 449

Snodgrass, C., Lowry, S. C., Fitzsimmons, A.. 2006, MNRAS, 373, 1590

Whipple, F. L. 1984, Icarus, 60, 522

Wilkening, L. L., Sherrod, P. C., Sekanina, Z. 2007, CBET, 1118

\clearpage 

\begin{deluxetable}{llllllll}
\tabletypesize{\scriptsize}
\tablecaption{Journal of Observations \label{imaging}}
\tablewidth{0pt}
\tablehead{
\colhead{UT Date} & \colhead{Telescope}  & \colhead{Camera}   & \colhead{Filters} & \colhead{r$_{H}$ [AU] \tablenotemark{1}} & \colhead{$\Delta$ [AU] \tablenotemark{2}} & \colhead{$\phi$ [deg] \tablenotemark{3}} & \colhead{Image scale [km/pixel]} }
\startdata
%2007 Oct 6 & UH 2.2m & Tek & R & 2.44 & 1.63 & 16.5 &  260\\
2007 Nov 6.4 & CFHT & MegaCam & r' & 2.49 & 1.62 & 13.7 & 218\\
2007 Nov 8.5 & CFHT & MegaCam & r' & 2.50 & 1.62 & 13.3 & 218\\
2007 Nov 9.5 & CFHT & MegaCam & r' & 2.50 & 1.62 & 13.1 & 218\\
2007 Nov 10.5 & CFHT & MegaCam & r' & 2.50 & 1.62 & 12.9 & 218\\
2007 Nov 11.5 & CFHT & MegaCam & r' & 2.51 & 1.62 & 12.7 & 218\\
2007 Nov 12.5 & CFHT & MegaCam & r' & 2.51 & 1.62 & 12.5 & 219\\
2007 Nov 13.5 & CFHT & MegaCam & r' & 2.52 & 1.63 & 12.3 & 219\\
2007 Nov 14.5 & CFHT & MegaCam & r' & 2.52 & 1.63 & 12.2 & 219\\
2007 Nov 15.5 & CFHT & MegaCam & r' & 2.52 & 1.63 & 12.0 & 219\\
\enddata
\tablenotetext{1}{Heliocentric distance}
\tablenotetext{2}{Geocentric distance}
\tablenotetext{3}{Phase Angle}
\label{table:obs}
\end{deluxetable}

\clearpage

\begin{deluxetable}{lllll}
\tabletypesize{\scriptsize}
\tablecaption{Fragment Characteristics \label{fragmags}}
\tablewidth{0pt}
\tablehead{
\colhead{Fragment } & \colhead{Magnitude  \tablenotemark{1} } & \multicolumn{2}{c}{Radius [m]}
%\colhead{SizeA} & \colhead{SizeB} 
& \colhead{Velocity \tablenotemark{2} [m s$^{-1}$]} \\ 
\multicolumn{2}{c}{} & \colhead{Model A \tablenotemark{3} } & \colhead{Model B \tablenotemark{4}} & \colhead{}
}
\startdata
A & 19.9 & 1009 & 37  & 104 $\pm$ 55 \\
B & 19.5 & 1234 & 45 & 65 $\pm$ 37\\
C & 23.5 & 194 & 10 & 56 $\pm$ 2 \\
D & 18.6 & 1850 & 68  & 35 $\pm$ 1\\
E &  18.3 & 2101 & 77 & 80 $\pm$ 12\\
F & 18.7 & 1785 & 66 & 48 $\pm$ 33\\
G & 20.4 & 820 & 30 & 44 $\pm$ 62\\
H & 18.3 & 2149 & 79 & 55 $\pm$ 34\\
I & 18.8 & 1690 & 62 & 108 $\pm$ 48\\
J & 18.8 & 1678 & 62 & 112 $\pm$ 78\\
K & 17.7 & 2853 & 105 & 123 $\pm$ 24\\
L & 17.6 & 2942 & 108 & 110 $\pm$ 68\\
M & 17.6 & 2991 & 110 & 91 $\pm$ 36\\
N & 18.9 & 1618 & 60 & 125 $\pm$ 53\\
O & 18.5 & 1895 & 70 & 88 $\pm$ 146\\ 
P & 20.4 & 791 & 29 & 102 $\pm$ 68\\
\enddata
\tablenotetext{1}{Determined using an aperture of radius $2^{\prime\prime}.22$ and from the image in which the fragment was first detected.}
\tablenotetext{2}{Velocity errors are obtained by resampling from the set of pairwise
day--to--day velocities, and computing the 68\% limit of the absolute
deviation from the un--resampled median.}
\tablenotetext{3}{Radius calculated assuming a geometric albedo of 0.1.}
\tablenotetext{4}{Radius calculated assuming activity similar to that of the nucleus.}
\label{table:megafrag}
\end{deluxetable}

\clearpage

\begin{deluxetable}{lll}
\tabletypesize{\scriptsize}
\tablecaption{Date of Closest Approach to Nucleus}
\tablewidth{0pt}
\tablehead{
\colhead{Fragment} & \colhead{UT of Closest Approach} & \colhead{Distance of Closest Approach [$^{\prime\prime}$]}}
\startdata
%\begin{table}
%\centering
 %\begin{tabular}{clc}
 %\\\hline
 %Fragment &  \multicolumn{1}{c}{UT of Closest} &  \multicolumn{1}{c}{Dist. of Closest}  \\
    %      &  \multicolumn{1}{c}{Approach}    &   \multicolumn{1}{c}{Approach ($^{\prime\prime}$)} \\
 %\hline\\
A  &   2007-10-27.2   &   14.8\\
B  &   2007-10-22.5   &   20.1\\
C  &   2007-10-22.9   &   15.6\\
D  &   2007-10-23.2   &   28.1\\
E  &   2007-10-22.7   &   16.6\\
F  &   2007-10-20.6   &   3.1\\
G  &   2007-10-22.0   &   27.8\\
H  &   2007-10-05.7   &   44.3\\
I  &   2007-10-24.7   &   37.8\\
J  &   2007-10-22.1   &    5.3\\
K  &   2007-10-29.9   &   15.5\\
L  &   2007-10-26.1   &   0.6\\
M  &   2007-10-24.4   &   16.2\\
N  &   2007-10-31.8   &   19.1\\
O  &   2007-10-25.7   &   13.6\\ 
P  &   2007-11-01.4   &   31.4 \\ %\hline
%\end{tabular}
\enddata
%\caption{\label{table:lumpapproach} The UT dates and distances of
  %closest approach to the nucleus for all of the fragments, computed by
%extrapolating each fragment's median measured position, and median pairwise
%velocity.}
%\end{table}
\label{table:lumpapproach}
\end{deluxetable}

\clearpage

% might need to insert caption to explain models presented below

\begin{deluxetable}{lll}
\tabletypesize{\scriptsize}
\tablecaption{Models of the Spatial Distribution of Fragments}
\tablewidth{0pt}
\tablehead{
\colhead{Fragment Distribution} & \colhead{Radial $p_{\rm KS}$} & \colhead{Velocity $p_{\rm KS}$}}
\startdata
Filled sphere & 0.54 & 0.31\\
Thin spherical shell & 0.0017 & 0.12\\
20\% spherical shell & 0.022 & 0.59\\
50\% spherical shell & 0.84 & 0.45\\
$r^{-1}$ density & 0.10 & 0.071\\
$r^{-2}$ density & 0.00029 & 0.003\\
Thin cone & 0.25 & 0.98\\
Solid cone & 0.05 & 0.016\\
Mixed thin and solid cone & 0.38 & 0.33\\ %\hline
    %\begin{table}
%\centering
  %\begin{tabular}{lllll}
    %\\\hline
%Fragment Distribution          & Radial $p_{\rm KS}$ &  Velocity $p_{\rm KS}$ \\
   % \hline\\
 % \end{tabular}
%\caption{\label{table:dismodels} Kolmogorov-Smirnov (KS) probabilities,
  %$p_{\rm KS}$, resulting from the comparison of the observed radial
  %and velocity fragment distributions with various models.     
  %A 20\% spherical shell is defined as a spherical distribution that is populated
  %from 0.8 to 1.0 times its maximum radius, and is unpopulated from 0.0 to 0.8.
  %A filled spherical  distribution, a 50\% thick shell, and a mixed cone model consisting
  %of an equal mixture of solid cone and thin cone agree best with the
  %data.  Only the thin spherical shells can be ruled out at the $2$ to
  %$3\sigma$ levels.  The radial and velocity data are not independent,
  %and in cases of disagreement the radial data are more reliable.}
%\end{table}
\enddata
\tablecomments{\label{table:dismodels} Statistical agreement (Kolmogornov-Smirnov
$p$) of various model fragment distributions with the distribution of
the fragments on the sky. The models are described in \S\ref{subsec:phasespace}.}
\end{deluxetable}

\clearpage

%\begin{deluxetable}{lll}
%\tabletypesize{\scriptsize}
%\tablecaption{Fragment Sizes [m]}
%\tablewidth{0pt}
%\tablehead{
%\colhead{Fragment} & \colhead{Model A \tablenotemark{a}} & \colhead{Model B} }
%\startdata
%A & 1009 & 37 \\
%B & 1234 & 45 \\
%C & 194 & 10 \\ 
%D & 1850 & 68 \\
%E & 2101 & 77 \\
%F & 1785 & 66 \\
%G & 820 & 30 \\
%H & 2149 & 79 \\
%I & 1690 & 62 \\
%J & 1678 & 62 \\
%K & 2853 & 105 \\
%L & 2942 & 108 \\
%M & 2991 & 110 \\
%N & 1618 & 60 \\
%O & 1895 & 70 \\
%P & 791 & 29 \\
%\enddata
%\tablenotetext{a}{Size calculated assuming a geometric albedo of 0.1 using the image in which the fragment was first detected.}
%\label{table:sizes}
%\end{deluxetable}

%\clearpage

%\begin{deluxetable}{ll}
%\tabletypesize{\scriptsize}
%\tablecaption{Fragment Velocities}
%\tablewidth{0pt}
%\tablehead{
%\colhead{Fragment} & \colhead{Velocity [m s$^{-1}$]}}
%\startdata
%A & 104 \\
%B & 65 \\
%C & 56 \\
%D & 35 \\
%E & 80 \\
%F & 48 \\
%G & 44 \\
%H & 55 \\
%I & 108 \\
%J & 112 \\ 
%K & 123 \\
%L & 110 \\
%M & 91 \\
%N & 125 \\
%O & 88 \\
%P & 102 \\
%\enddata
%\label{table:vel}
%\end{deluxetable}

%\clearpage

\clearpage

\clearpage

\clearpage

\clearpage

\clearpage

\clearpage

\clearpage

\clearpage

% need to insert new figure

\clearpage

% need to fix figure

\clearpage

\clearpage

%\begin{figure}
  %\centering 
  %\includegraphics[angle=-90,  totalheight=7cm]{lump-flux-vs-dlogfluxdt.ps} 
  %\caption{\label{fig:lump-flux-vs-dlogfluxdt} Fading rate as a
    %function of fragment brightness. The fading rate is shown for each
    %fragment over the first three nights of data.  Six points lie off
    %the plot.  The dotted (solid) line shows the median fading rate,
    %including outliers, for Flux $>5\times 10^4$ (Flux $<5\times 10^4$).  
    %There is no apparent dependence of the fading rate on fragment flux.
   % }
%\end{figure}

\clearpage

\begin{figure}
\centering 
\includegraphics[angle=-90, totalheight=10cm]{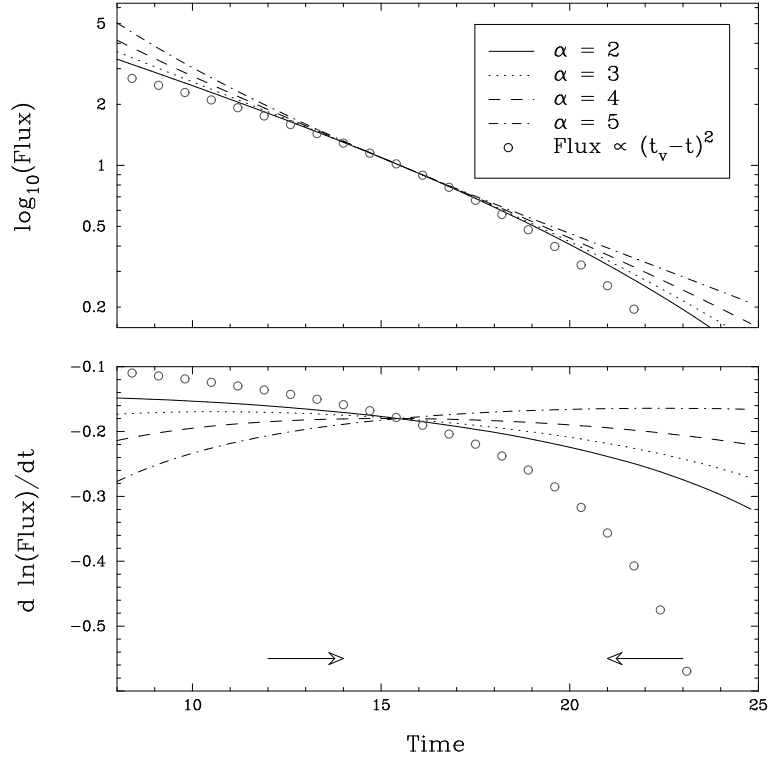} 
%\plotone{lumps.ps}
  \caption{\label{fig:powerlaw-flux-fallofs} The fragment flux and its
    logarithmic derivative as a function of time, assuming that each
    fragment actually consists of a ${\ell_{\rm f}}^{-\alpha}$
    differential distribution of particles, and that each particle has
    a flux proportional to its area ${\ell_{\rm f}}^2$, as desribed in
    equations \ref{eq:plaw1} to \ref{eq:plaw3}.  At each $\alpha$, we
    adjust the one free parameter so that $\dot F/F =0.2$ at $t=15.5$,
    as seen in Figure \ref{fig:lump-r-vs-flux}.  It is evident that
    all of the exponents $\alpha$ produce an approximately exponential
    decay curve during the observation timespan, denoted by the the
    arrows. In contrast, a single-fragment model ($\circ$ symbols) with
    $F(t)\propto (t_v-t)^2$ produces an accelerating falloff during
    the span of observations.  }
\end{figure}

\appendix

\section{Constant fractional fading per unit time: a swarm of particles?}

As noted above, there are several problems with a model consisting of
monolithic sublimating fragments of material whose brightness scales as
the surface area.  The brightest (largest) fragments are not the
closest to the nucleus, the least bright fragments do not fade notably
faster, and the best fit fading law is exponential rather than
quadratic in time.

Accordingly, we consider ``fragments'' that are in fact swarms of
particles obeying a power law distribution instead of
single large fragments.   Such a power law distribution arises naturally
in collisional fragmentation or grinding processes (Dohnanyi 1969).
If the fragments are actually a distribution of small particles rather than monolithic pieces,
their brightness, fading rate, and radial distribution are no longer expected to
be coupled,  which is what we observe.

\noindent We assume that each fragment consists of a differential distribution of 
sub--fragments of size ${\ell_{\rm f}}$:

\begin{equation}
\label{eq:plaw1}
 N(\ell_{\rm f}) \propto \ell_{\rm f}^{-\alpha} 
 \quad{\rm for}\quad 
 \ell_{\rm f} \in [{\ell_{\rm f}}_1, {\ell_{\rm f}}_2]
\end{equation}

\noindent Taking a constant sublimation rate $\dot \ell_{\rm f}<0$ summed over
the combined area of the sub--fragments, this distribution produces an observed flux:

\begin{equation}
\label{eq:plaw3}
 F(t,\alpha,{{\ell_{\rm f}}_1},{{\ell_{\rm f}}_2})
\propto 
 \int_{\max({\ell_{\rm f}}_1-\dot \ell_{\rm f} t,0)}^{\max({\ell_{\rm f}}_2 - \dot \ell_{\rm f} t),0)} 
 \ell_{\rm f}^2 (\ell_{\rm f} - \dot \ell_{\rm f} t) ^{-\alpha} \, d \ell_{\rm f} 
\end{equation}

\noindent 

${\dot F}(t)/F(t)$ is readily shown to be inversely proportional to
$t_{\rm v}={{\ell_{\rm f}}_2}/{\dot\ell}$, the timescale of the
vanishing of the largest fragment ${{\ell_{\rm f}}_2}$.
In Figure \ref{fig:powerlaw-flux-fallofs}, we consider
several values of the exponent $\alpha$, and for each value we fix
value of $t_{\rm v}$ that gives $d/dt \ln F(t) =-0.18$ at t=15.5, like
the real data.   We find that for all $\alpha$ considered, ${\dot F}(t)/F(t)$
is constant over the observational window, in agreement with
our fit of the actual fading. 

We conclude that fragments are plausibly explained as clusters
of sub-fragments, obeying a power law distribution, with the power law index 
anywhere between $-2$ and $-4$.  Such a model is consistent with the observed constant 
logarithmic fading rate, and allows the fading rate, fragment brightness, and distance from the 
nucleus to be independent, as observed.

\end{document}